\def\year{2018}\relax
\documentclass[letterpaper]{article} 
\usepackage{aaai18}  
\usepackage{times}  
\usepackage{helvet}  
\usepackage{courier}  
\usepackage{tabularx}
\usepackage{url}  
\usepackage{graphicx}  
\begin{document}
%
\title{Mining the Relationship between Emoji Usage Patterns and Personality}
\author{Weijian Li*, Yuxiao Chen*, Tianran Hu, Jiebo Luo\\
University of Rochester\\
Rochester, NY 14623\\
\{wli69, ychen211, thu, jluo@cs.rochester.edu\}\\
}

\maketitle
\begin{abstract}
Emojis have been widely used in textual communications as a new way to convey nonverbal cues. An interesting observation is the various emoji usage patterns among different users.  In this paper, we investigate the correlation between user personality traits and their emoji usage patterns, particularly on overall amounts and specific preferences. To achieve this goal, we build a large Twitter dataset which includes 352,245 users and over 1.13 billion tweets associated with calculated personality traits and emoji usage patterns. Our correlation and emoji prediction results provide insights into the power of diverse personalities that lead to varies emoji usage patterns as well as its potential in emoji recommendation tasks.
\end{abstract}

\section{Introduction}
Emojis are a large set of pictorial representations \cite{miller2016blissfully} and are widely used for textual Computer-Mediated Communications (CMC). Researchers find that the same Emoji is used and interpreted differently by different users \cite{miller2017understanding}. Several studies \cite{hu2017spice,ai2017untangling} have been done to find the reason behind this observation. Other studies \cite{felbo2017using,barbieri2017emojis} focus on real applications, shows the potential of predicting emojis given user text information. Their methods mainly based on emoji context meanings, sentiments, and linguistic effects and less focus on the differences of human attributes. Inspired by studies that extract divergent human personalities from social media data \cite{liu2016analyzing,wei2017beyond} and assessing participants\rq{} personalities from emojis \cite{marengo2017assessing} in a relatively small scale, we consider user personality may be one of the factors that lead to different emoji usage patterns and may also contribute to emoji recommendation tasks.

To this end, our paper dives deeper to explore the hidden relationship between user personality and emoji usage patterns from a data-driven and fully computational perspective. By doing this, we are able to extract personality traits and emoji usage patterns from collected volumetric tweets data to avoid influences from subjective surveys or questionnaire answers.
    
In summary, our work makes two main contributions:
\begin{itemize}
\item We show personality as an influential factor in different emoji usage patterns by presenting a fully computational analysis using large-scale social media data. 
\item We present personality\rq{}s discriminant power in the emoji prediction task with a superior prediction result when personality is fused as additional input features into the machine learning classifier.
\end{itemize}

\section{Data Collection and Preprocessing}
Our dataset contains 352,245 twitter users collected from March 2016 to June 2016. For each user, we collect his/her most recent 3,200 tweets until June 2016 which gives us 1.13 billion tweets in total. Since we are more interested in individual users instead of public accounts whose tweet context may not relate to the account owner, we keep users whose number of followers less than 1649 (90th percentile of follower distribution) and the number of followees less than 1180 (90th percentile of followee distribution).  Secondly, for each remained user, we remove the tweets that are: not written in English, contain less than ten words after removing hashtags, URLs, mentions and email address. Users who have less than 500 tweets left are removed to keep enough context information for our later analysis. Lastly, to conduct our research with a focus on the majority of users, we filter users who frequently or seldom use emojis. To be more specific, for each user, his/her emoji usage frequency is defined as the number of tweets containing emojis divided by the total number of tweets. Users whose emoji usage frequency is more than 0.95 or less than 0.05 are filtered out. Eventually, we end up with 71,823 users, 75 million tweets.

\section{Extraction of User Personality Traits}
In this paper, we use Big Five personality traits (Big Five) \cite{john1999big} to describe user personality. Big Five is one of the popular and widely adopted definitions of personality. It models human personality as five different dimensions: Openness (degree of appreciation to new, unusual ideas and experiences), Conscientiousness (the tendency to self-discipline), Extraversion (the extent of preference to being in social situations
), Agreeableness (willingness to cooperate or compassionate) and Neuroticism (level of emotional stability). We use abbreviations Open., Cons., Extra., Agree., Neuro., or O, C, E, A, N for each dimension if necessary. Each user\rq{}s Big Five is computed based on their tweet text data leveraging a LIWC \cite{pennebaker2015development} based personality calculation model Receptiviti\footnote{https://www.receptiviti.ai/, retrieved Jan. 2018.} which is suggested in the LIWC paper. We take each user\rq{}s whole processed tweets as input and obtain each user\rq{}s psychological attributes from Receptiviti. We keep the Big Five personality scores from the attributes as personality traits for the users.

\section{Emoji Usage Analysis}
In this section, we reveal the relationship between users\rq{} personalties and their mean emoji usage frequencies, as well as the relationship with emoji usage preferences. 

\subsection{Usage Frequency Analysis}
\begin{table}
\begin{center}
	\begin{tabularx}{\linewidth}{XXXXXXXX}
\hline
Dimension& P-mean&N-mean. &t-value & p-value.\\\hline
Open. & 0.237 & 0.233 & 2.147 &{\bf 0.032}\\
Cons. &0.229 &{\bf 0.238} &-5.001 &5.7e-7\\
Extra. & 0.225 & {\bf 0.273}&-25.58&4.14e-43\\
Agree.&{\bf 0.261}&0.237&12.69&7.67e-37\\
Neuro.&{\bf 0.258}&0.223&19.34&6.76e-83\\\hline
\end{tabularx}
\end{center}
  \caption{The result of t-test on mean emoji usage frequency between positive (P) and negative (N) sets in each five personality dimensions. The last column represents two-tailed p-values. All results are significant (two-tailed p-value $<$ 0.005) except openness.
}
  \label{tbl:table1}
  
\end{table}
Here we are interested in how each dimension of the five personalities affects mean emoji usage frequency. For each personality dimension, we collect representative users who have the top and bottom quartile personality scores, mark them as the positive set and the negative set respectively. By doing this, we obtain two sets of users on each personality dimension which gives us ten sets of users in all. For each set in the pair sets (positive and negative), we compute mean emoji usage frequencies, as defined in the previous section, for the users within the set, and conduct two-tailed t-test over two sets to verify the significance of differences for mean usage frequencies. Our results are reported in Table 1.

\begin{itemize}
\item {\bf Openness}  It is obvious to see four out of five results are significant at two-tailed p-value $<$ 0.005, except for openness. For openness, since its p-value is greater than the significant level, we cannot reject the null hypothesis on the equality of two mean values. We consider the reason is that emojis are predefined pictorial representations and have been used for online communications for a long time. Whether to use emojis or not does not depend on the willingness of accepting new things. Thus, we infer that users with different openness scores use a similar amount of emojis in their tweets.
    
\item{\bf Conscientiousness}  Users with a higher score in conscientiousness use less emojis than users with lower conscientiousness scores. This is because of their vigilance and self-discipline which make them stick to traditional and plainer text contexts to conserve their emotions.

\item{\bf Extraversion}  Users with a low score in extraversion use much more emojis than users with higher scores. What\rq{}s more, they are the users who use emojis the most. We consider the reason is that introversion users prefer implicit visual contexts than explicit text messages on which sender has to express themselves directly.

\item{\bf Agreeableness}  Users with high agreeableness score use more emojis than users with lower scores. By definition, high agreeableness score indicates distinct attributes like empathetic and compliance. Messages convey these ideas are not easy to describe directly using text words. Instead, they are more expressible via vivid facial or object representations like emojis.

\item{\bf Neuroticism}  Users with a high score in neuroticism have unstable emotions and have difficulty to control their emotions well. Given their unpredictable emotions, we consider neuroticism users use more emojis than plain text words to better describe their emotional feelings.
 \end{itemize}

\subsection{Usage Preference Analysis}
\begin{table*}
\centering
	\includegraphics[width=\linewidth,height=8cm]{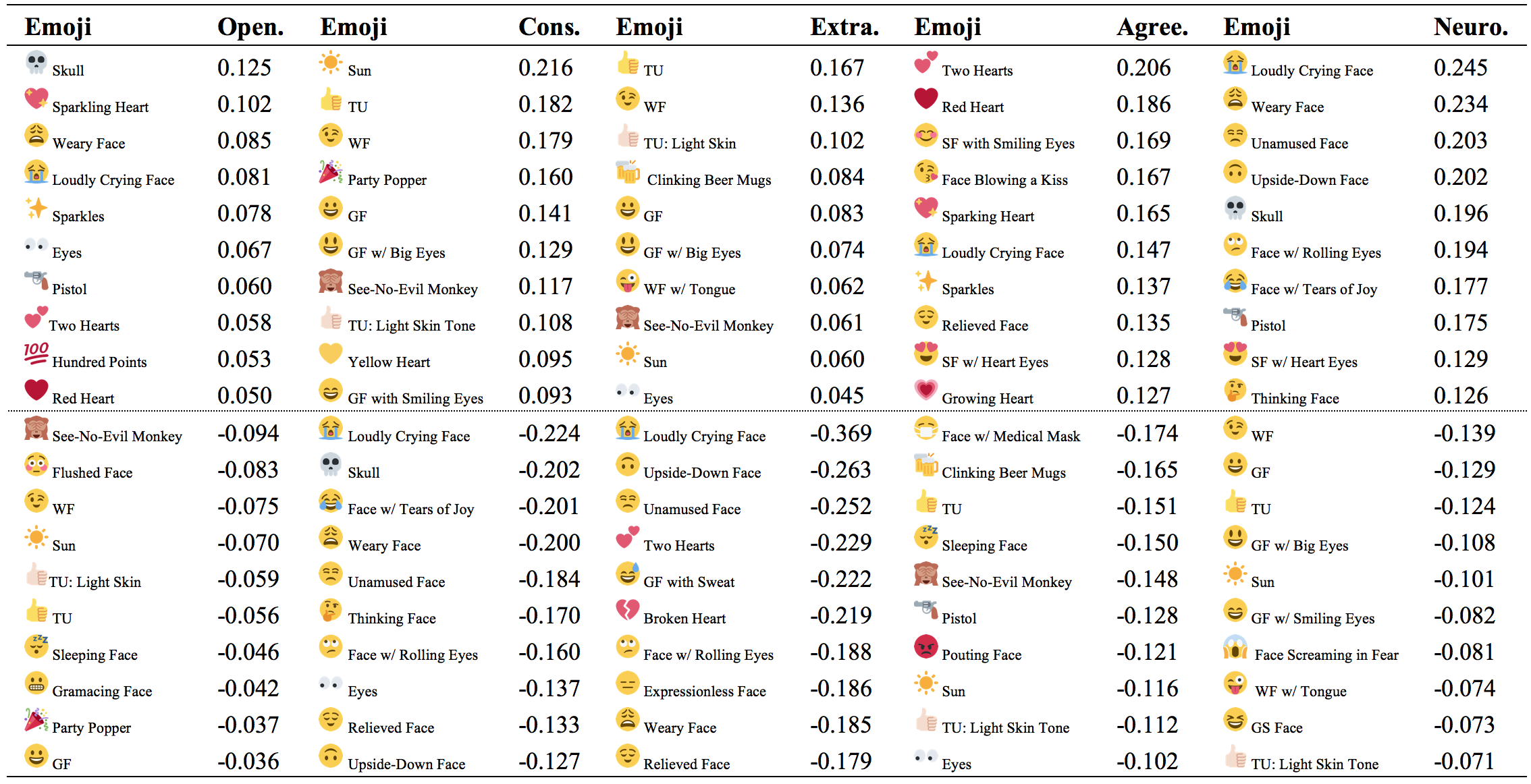}
  \caption{Top 10 positive and negative Spearman Correlation Coefficient Results on each Big Five Personality dimension. All coefficient scores significant at Bonferroni-corrected p $<$ 0.005. Correlation results on each dimension are listed on the right columns. GF: Grinning Face, WF: Winking Face, GS: Grinning Squinting, TU: Thumbs Up.
}
  \label{tbl:table2}
\end{table*}
Our previous analysis measures mean emoji usage frequencies and shows how emoji usage frequency varies along each of five personality dimensions. To extend our study quantitatively and more specifically, we investigate which emojis are more preferred on five personality dimensions. By the time we conducted our experiment, there were in total 2685 emojis\footnote{The emojis used in this study are extracted from Twemoji v2.4.
https://emojipedia.org/twitter/, retrieved Jan. 2018.
} in the dataset. Here we focus on the most frequently used emojis based on the overall usage frequency. To obtain these emojis, we first collect overall emoji frequencies over the entire dataset and sort the popularity of emojis in descending order. Emojis whose frequency is larger than 0.1\% of total emoji frequency are selected as our study objects. Following these steps, we obtain 67 most frequently used emojis. We compute Spearman’s Rank Correlation Coefficient (Kendall 1938) between five personality traits and their emoji usage frequencies. Our results are reported in Table 2. For each personality dimension, we list the top and bottom ten correlated emojis on the left column and corresponding correlation coefficient scores on the right column. 
  
{\bf Openness}	In general, correlation coefficient scores in openness are relatively small in scale indicating similar usage preferences on each emoji. This result coincides with our previous findings on emoji usage frequency. Besides, there are no clear emoji preference patterns based on the definition of openness. Thus, we infer that openness has little correlation with emoji usage patterns.
  
{\bf Conscientiousness}	 People with higher conscientiousness scores are more vigilant and self-controlled. They use less negative emotion emojis such as  
\includegraphics[height=0.8em]{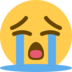} (-0.224),  \includegraphics[height=0.8em]{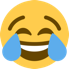} \cite{wijeratne2016emojinet} (-0.201) ,  \includegraphics[height=0.8em]{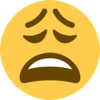} (-0.200) and  \includegraphics[height=0.8em]{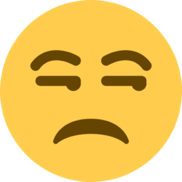} (-0.184). Instead, they prefer expressing positive emotions through emojis like  \includegraphics[height=0.8em]{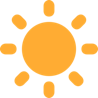} (0.216),  \includegraphics[height=0.8em]{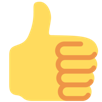} (0.182)  \includegraphics[height=0.8em]{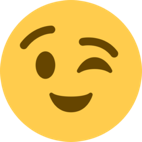} (0.179) and  \includegraphics[height=0.8em]{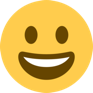} (0.141).  \includegraphics[height=0.8em]{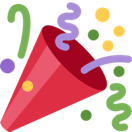} (0.160) is also used frequently. Though this emoji is explained as \lq\lq{}Party Popper\rq\rq{}, we find people use it regularly when sending congratulations and celebrating achievements which are precisely positive emotional circumstances.

{\bf Extraversion}	Similar to conscientiousness, we find people high in extraversion dimension prefer positive emotion emojis such as \includegraphics[height=0.8em]{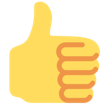}(0.167),  
\includegraphics[height=0.8em]{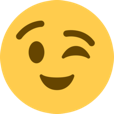} (0.136) and  
\includegraphics[height=0.8em]{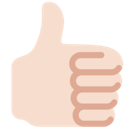} (0.102). On the other hand, they seldom use emojis like  
\includegraphics[height=0.8em]{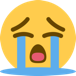} (-0.369),  
\includegraphics[height=0.8em]{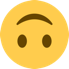} (-0.263) and  
\includegraphics[height=0.8em]{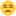} (-0.252) which represent negative or ambiguous facial expressions. 
We think \includegraphics[height=0.8em]{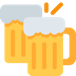} (0.084) is often used at celebrations or parties where many people are gathered together. This conforms their extravert personality.

{\bf Agreeableness}	We can see many heart shape (or contain heart shape) emojis such as  
\includegraphics[height=0.8em]{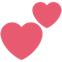} (0.206),  
\includegraphics[height=0.8em]{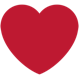} (0.186),  
\includegraphics[height=0.8em]{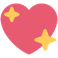} (0.165) and particularly  
\includegraphics[height=0.8em]{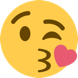} (0.167) are positively correlated with Agreeableness. We consider people are showing \lq\lq{}like\rq\rq{}, \lq\lq{}love\rq\rq{} and other empathetic meanings with strong positive emotions and inner feelings which are highly representative for users with high agreeableness scores. On the opposite, \includegraphics[height=0.8em]{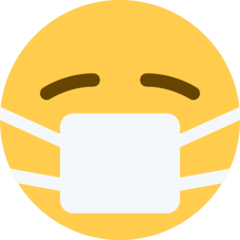} (-0.174),   
\includegraphics[height=0.8em]{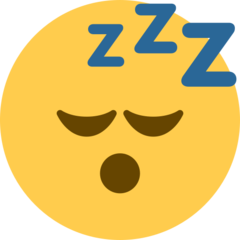} (-0.150) and 
\includegraphics[height=0.8em]{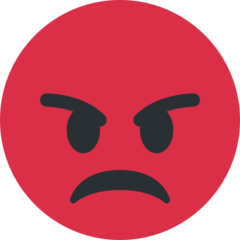} (-0.121) are seldom used because of their negative and “dislike” emotions.

{\bf Neuroticism}	It is interesting to see that neuroticism people prefer to use exaggerated facial expressions with rich emotions. For example, \lq\lq{}Loudly Crying Face\rq\rq{}  
\includegraphics[height=0.8em]{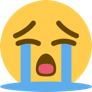} (0.245), \lq\lq{}Weary Face\rq\rq{} 
\includegraphics[height=0.8em]{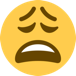} (0.234), \lq\lq{}Upside Down Face\rq\rq{}  
\includegraphics[height=0.8em]{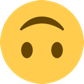} (0.202) and \lq\lq{}Face w/ Rolling Eyes\rq\rq{} 
\includegraphics[height=0.8em]{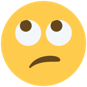} (0.194). These emojis have little positive correlation with other personality traits which represents a unique emoji usage pattern for neuroticism users as well as their distinct emotional characteristics.

\section{Predicting Emoji Usages}
\begin{table*}[ht]
\begin{tabularx}{\linewidth}{l|c|c|c|c|c}
\hline
     & \textbf{Acc} & \textbf{Tweet Text} & \textbf{GD} & \textbf{Pred. 1} & \textbf{Pred. 2} \\
\hline
\textbf{Text} & 0.158 & - & - & - & -\\
\hline
\textbf{Text + O} & 0.160 & - & - & - & -\\
\hline
\textbf{Text + C} & 0.166 & Stay positive mate, know how you feel, everyday is a new adventure & \includegraphics[height=1em]{e1.png} & \includegraphics[height=1em]{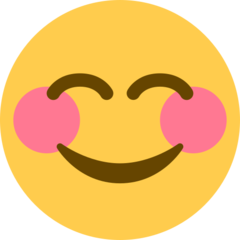} & \includegraphics[height=1em]{e1.png} \\
\hline
\textbf{Text + E} & 0.165 & Yay! I can't be in phoenix tonight, but can watch online & \includegraphics[height=1em]{e1.png} & \includegraphics[height=1em]{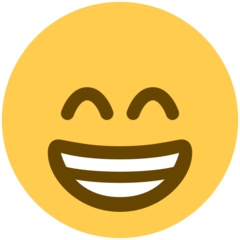} & \includegraphics[height=1em]{e1.png} \\
\hline
\textbf{Text + A} & 0.162 & Going to post an entry to celebrate the success of the album gone gold & \includegraphics[height=1em]{a2.png} & \includegraphics[height=1em]{e1.png} & \includegraphics[height=1em]{a2.png}\\
\hline
\textbf{Text + N} & 0.163 & I'm working 10:30 - 9:00 today ... how did I get myself into & \includegraphics[height=1em]{e5.png} & \includegraphics[height=1em]{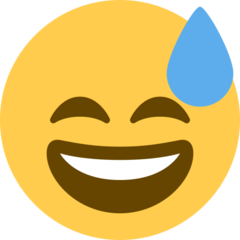} & \includegraphics[height=1em]{e5.png}\\
\hline
\textbf{Text + All} & \textbf{0.172} & - & - & - & - \\
\hline

\end{tabularx}
\label{acc}
\caption{Accuracy of test data set for each model. Text: using text feature only. Text + \lq{}X\rq{}: using text feature and one of five personality dimensions. Text + All:  using text feature and all 5 personality dimensions. Acc: Accuracy. GD: Ground Truth Label. Pred. 1: Prediction result using only text feature. Pred. 2: Prediction result using text feature and corresponding row\rq{}s personality trait.}
\end{table*}
To validate our observations under a real application, we build a classifier to predict 19 most frequently used emojis based on users\rq{} tweets texts data. Our goal is to see if fusing five personality traits as additional features would improve the classifier\rq{}’s performance.

Our final prediction results are reported in table 3. Given that our test set is balanced, we use accuracy here for intuitive representation. It is clear to see that the prediction performance is improved when the classifier is fused with personality data. Each personality dimension increases prediction result. Openness, which is found to be hardly correlated with emoji usage patterns, has a little improvement. Conscientiousness, on the other hand, provides the most information as a personality trait. Our best prediction result is obtained when all five personalities are given to the classifier which indicates the combined power of five personalities. The improved prediction results not only validate our previous correlation findings but also show the potential of personality feature in personalized emoji recommendation systems. 

To show the effects of personality traits, we report a sample result for each personality dimension in Table 3. The emojis are initially  classified incorrectly by only using text information, but are correctly classified after fusing the corresponding row\rq{}s personality information. Concretely, for conscientiousness, extraversion, and neuroticism, the classifier correctly labeled the sample tweet with emojis in correct sentiments but fails to predict the exact emoji type for the user. For agreeableness, the classifier gives a reasonable prediction based on words “success” and “celebrate”, while the user who posted this tweet prefer to use  \includegraphics[height=0.8em]{a2.png}. After fusing personality traits, our classifier correctly labels all these examples by learning the personalized information from personality.

\section{Discussion and Conclusion}
In this paper, we investigate the hidden relationship between user personality and emoji usage patterns in Twitter tweets using a computational approach. Generally speaking, users with low extraversion scores use emojis most frequently while users with low neuroticism scores use emojis the least often. From another perspective, we find high correlation values and specific emoji usage in line with perceived user\rq{}s personality traits. However, some exceptions do exist. For example, Openness, as a sign of the wiliness to new objects, shows no obvious relationship with emoji usage. We consider emoji, itself as a pre-designed form of non-verbal communication tool, may suggest little information on openness characteristics. Finally, our prediction results show discriminant power of personalities in differentiating different emoji usage patterns. Best prediction performance is achieved by fusing all five personalities into the developed classifier. We believe our findings can benefit researchers and engineers in designing personalized emoji recommendation systems.
    
Future work can be directed at studying personality influences under more specific circumstances such as under different topics. Although personality is a human characteristic, users\rq{} personality may change on different topics.

\section{Acknowledgement}
We thank the support of New York State through the Goergen Institute of Data Science.

\bibliography{Bibliography-File}
\bibliographystyle{aaai}
\end{document}